\documentclass[aps,preprint]{revtex4}%
\usepackage{amsfonts}
\usepackage{amsmath}
\usepackage{amssymb}
\usepackage{graphicx}%
\begin{document}
\title{Entanglement in nuclear quadrupole resonance }
\author{G. B. Furman$^{1,2}$, V. M. Meerovich$^{1}$, and V. L. Sokolovsky$^{1}$}
\affiliation{$^{1}$Physics Department, Ben Gurion University, POB 653, Beer Sheva, 84105, Israel}
\affiliation{$^{2}$Ohalo College, POB 222, Katsrin, 12900, Israel}

\begin{abstract}
Entangled quantum states are an important element of quantum information
techniques. We determine the requirements for states of quadrupolar nuclei
with spins
$>$%
1/2 to be entangled. It was shown that entanglement is achieved at low
temperature by applying a magnetic field to a quadrupolar nuclei possess
quadrupole moments, which interacts with the electricfield gradient produced
by the charge distribution in their surroundings.

\end{abstract}
\maketitle

\section{\textbf{Introduction}}

Quantum entanglement \cite{Benenti books 2007,Amico 2008 rev,HorodeckiR 2009
rev} , the most characteristic feature of quantum mechanics, is one of the
central concepts in quantum information theory and is the feature that
distinguishes it most significantly from the classical theory. Entanglement is
now viewed as a physical resource, which provides a means to perform quantum
computation and quantum communication \cite{Bennett CH 1984} . It should be
emphasized that property of entanglement can be considered regardless of the
nature of qubits\cite{Barnum,Can2007}.

One of the most intensively investigated systems is clusters of coupling
nuclear spins \cite{Benenti books 2007,Amico 2008 rev,HorodeckiR 2009 rev}
which received considerable attention as a platform for the practical
implementation of a quantum computer (QC) by using nuclear magnetic resonance
(NMR) technique \cite{Gershenfeld1997,Gory DG1997,Jones2000,Marjanska2000}.
The strength of the coupling (such as dipole-dipole, scalar or exchange
interactions) between different spins and interaction of the spins with
environment determine the decoherence time. Short decoherence times limit
possible calculation times of the QC \cite{ASteane1998}. Later it has been
proposed to used quadrupole nuclei, thus eliminating the requirement of
interaction between spins and decreasing interaction of qubits with
environment \cite{ARKessel1999,ARKessel 2000,A
KKhitrin2000,AKKhitrin2001,Furman2002,GBFurman2002,Zobov2007} \ 

It was shown that a spin $\frac{{\LARGE 3}}{{\LARGE 2}}$ system is equivalent
to a system of two magnetization vectors \cite{MBloom1955,TPDas1957} and can
be represented using the Pauli spin matrices 2x2 \cite{GWLeppermeier1966}. It
means that a single spin 3/2 is isomorphic to a system consists of two dipolar
coupling spins 1/2, which can be considered as qubits. This was experimentally
confirmed by using NQR technique with quadrupole splitting \cite{A
KKhitrin2000,AKKhitrin2001}. The feasibility of quantum computing based on a
pure (without external magnetic fields) NQR technique was theoretically
investigated in detail in \cite{Furman2002}. Using the resonance excitation
technique and the level-crossing method, it was proposed three quantum logic
gates:\emph{ }a controlled $\ NOT$, $SWAP$ and $NOT_{2}$ . Thus the method to
synthesize qubits from a set of the spin states of a single particle with spin
higher 1/2 has been developed. It is logical to raise the question of
entanglement of these qubits.

Our present purpose is to investigate entanglement between the quantum states
of quadrupole nuclei. We consider a nucleus with spin $\frac{3}{2}$ being in
an internal electric field gradient (EFG) and an external magnetic field when
the quadrupole interaction energy is of the order of the magnitude or even
greater than the Zeeman one. Results of computer simulations of entanglement
dynamics are presented for real spin systems with a non-equadistant energy spectrum.

\section{\textbf{NQR in magnetic field}}

Let us consider a quadrupole nucleus with spin $I$ ($I>1/2$) in the
thermodynamic equilibrium with the density matrix
\begin{equation}
\rho=%
\mathbb{Z}
^{-1}\exp\left(  -\frac{\mathcal{H}}{k_{B}T}\right)  . \tag{1}%
\end{equation}
Here $k_{B}$ is the Boltzmann constant, $T$ is the lattice temperature, $%
\mathbb{Z}
=Tr\left\{  \exp\left(  -\frac{\mathcal{H}}{k_{B}T}\right)  \right\}  $ is the
partition function. In the general case, the Hamiltonian $\mathcal{H}$ can
consist of the Zeeman ($\mathcal{H}_{M}$) and the quadrupole ($\mathcal{H}%
_{Q}$) parts,%
\begin{equation}
\mathcal{H}=\mathcal{H}_{M}+\mathcal{H}_{Q}\text{.} \tag{2}%
\end{equation}
The Zeeman interaction between the applied magnetic field, $\vec{H}_{0}$ and
nuclear spin is the external factor for a crystal and the direction of this
field is chosen as the $z$-axis of the laboratory frame, $\vec{H}_{0}%
=H_{0}\vec{z}$, where $H_{0}$ is the strength of the external magnetic field.
The part of the Hamiltonian describing this interaction is
\begin{equation}
\mathcal{H}_{M}=-\gamma H_{0}I_{z}, \tag{3}%
\end{equation}
where $\gamma$ is the gyromagnetic ratio of the nucleus with spin\textit{ }%
$I$, $I_{z}$ is the projection of the individual spin angular momentum
operators $\vec{I}$ on the $z$- axis.

Quadrupole coupling exists between a non-spherical nuclear charge distribution
and an electric field gradient (EFG) generated by the charge distribution in
their surroundings. It is possible to reduce the EFG symmetric tensor to a
diagonal form by finding the principal axes frame (PAF) with the $Z$- and
$X$-axises directed along the maximum and minimum of EFG, respectively,
$\left\vert V_{ZZ}\right\vert \geq\left\vert V_{YY}\right\vert \geq\left\vert
V_{XX}\right\vert $, where $V_{\xi\xi}=\frac{\partial^{2}V}{\partial\xi^{2}%
}\ \left(  \xi=X,Y,Z\right)  $ and $V$ is the potential of the electric field.
In the laboratory frame the quadrupolar Hamiltonian can be presented in the
following form%

\begin{equation}
H_{Q}=\frac{eQq_{ZZ}}{4I(2I-1)}U\left(  \theta\ ,\varphi\right)  \left[
3I_{z}^{2}-\vec{I}^{2}+\frac{\eta}{2}\left(  I_{+}^{2}-I_{-}^{2}\right)
\right]  U^{+}\left(  \theta\ ,\varphi\right)  ,\tag{4}%
\end{equation}
where
\begin{equation}
U\left(  \theta\ ,\varphi\right)  =e^{-i\varphi Iz}e^{-i\theta Iy}e^{i\varphi
Iz},\tag{5}%
\end{equation}
$eQq_{ZZ}$ is the quadrupole coupling constant of EFG, $I_{\pm}$ are the
raising and lowering operators of the spin, and $\theta\ $and $\varphi$ refer
to the polar and azimuthal angles determining the orientation of the
laboratory frame $z$-axis\emph{ }in the PAF coordinate system. The asymmetry
parameter\emph{ }$\eta$ is defined as
\begin{equation}
\eta=\frac{V_{YY}-V_{XX}}{V_{ZZ}},\tag{6}%
\end{equation}
which may vary between $0$ and $1$.

Using the Hamiltonian (2) the density matrix (1) can be represented as a
function of parameters $\alpha=\frac{\gamma H_{0}}{k_{B}T}$ and $\beta
=\frac{eQq_{ZZ}}{4I(2I-1)k_{B}T}$:%
\begin{equation}
\rho=%
\mathbb{Z}
^{-1}\exp\left\{  -\alpha I_{z}-\beta U\left(  \theta\ ,\varphi\right)
\left[  3I_{z}^{2}-\vec{I}^{2}+\frac{\eta}{2}\left(  I_{+}^{2}-I_{-}%
^{2}\right)  \right]  U^{+}\left(  \theta\ ,\varphi\right)  \right\}  .
\tag{7}%
\end{equation}
The density matrix (7) can be employed to obtain information on the dependence
of entanglement on the magnetic field, quadrupole coupling constant,
orientation of the crystal principal axis in the laboratory frame, and
temperature. \emph{ }

Below we consider entanglement in a system of spins 3/2.\emph{ }A suitable
system for studying by NQR technique is a high temperature
superconductor\ $YBa_{2}Cu_{3}O_{7-\delta}$ containing the $^{63}Cu$ and
$^{65}Cu$ nuclei with spin $\frac{3}{2}$ possessing quadrupole moments
$Q=-0.211\times10^{-24}$ cm$^{2}$ and $-0.195\times10^{-24}$ cm$^{2}$,
respectively \cite{MMali}. There are two different locations of copper ions in
this structure: the first are the copper ion sites at the center of an oxygen
rhombus-like plane while the second one is five-coordinated by an apically
elongated rhombic pyramid. The four-coordinated copper ion site, EFG is highly
asymmetric $\left(  \eta\geq0.92\right)  $ while the five-coordinated copper
ion site, EFG is nearly axially symmetric $\left(  \eta=0.14\right)  $. The
quadrupole coupling constant $\left(  eQq_{ZZ}\right)  $ of $^{63}Cu$ in the
four-coordinated copper ion site is $38.2$ MHz$\ \ $\ and in the
five-coordinated copper ion site is $62.8$ MHz \cite{MMali}.

\section{Reduced density matrix and measures of entanglement}

An important measure of entanglement is the concurrence \cite{Wootters1998}.
The concurrence $C$ is usually used \cite{Wootters1998}. For the maximally
entangled states, the concurrence is $C=1$, while for the separable states
$C=0$. The concurrence of a quantum system with the density matrix presented
in the Hilbert space as a matrix $4\times4$ is expressed by the formula
\cite{Wootters1998}:%
\begin{equation}
C=\max\left\{  0,2\nu-\sum_{i=1}^{4}\nu_{i}\right\}  \tag{8}%
\end{equation}
where $\nu=\max\left\{  \nu_{1},\nu_{2},\nu_{3},\nu_{4}\right\}  $ and
$\nu_{i}$ $\left(  i=1,2,3,4\right)  $ are the square roots of the eigenvalues
of the product
\begin{equation}
R=\rho_{red}\tilde{\rho}_{red},\tag{9}%
\end{equation}
where $\rho_{red}=Tr_{partial}\left(  \rho\right)  $ is the reduced density
matrix, $Tr_{partial}\left(  ...\right)  $ is the partial trace, and
\begin{equation}
\tilde{\rho}_{red}=G\bar{\rho}_{red}G\tag{10}%
\end{equation}
where $\bar{\rho}_{red}\left(  \alpha,\beta\right)  $ is the complex
conjugation of the reduced density matrix $\rho_{red}$ \ and
\begin{equation}
G=%
\begin{pmatrix}
0 & 0 & 0 & -1\\
0 & 0 & 1 & 0\\
0 & 1 & 0 & 0\\
-1 & 0 & 0 & 0
\end{pmatrix}
.\tag{11}%
\end{equation}

An another important measure of entanglement is the entanglement entropy. In
order to implement this measure of entanglement we first map the Hilbert space
of a spin 3/2 which is four dimensional onto the Hilbert space of two spins
1/2, $S_{A}$ and $S_{B}$ \cite{Goldman1992}. The entropy of entanglement is
defined as \cite{Bennett1996,Popascu}:%
\begin{equation}
E=-Tr\left(  \rho_{A}\log_{2}\rho_{A}\right)  =-Tr\left(  \rho_{B}\log_{2}%
\rho_{B}\right)  .\tag{12}%
\end{equation}
Here $\rho_{A}$ is the partial trace of $\rho_{A}=Tr_{B}\left(  \rho\right)  $
over subsystem $B$, and $\rho_{B}$ has a similar meaning.

This entropy is related to the concurrence $C$ by the following equation
\cite{Wootters1998}
\begin{equation}
E\left(  x\right)  =-x\log_{2}x-\left(  1-x\right)  \log_{2}\left(
1-x\right)  ,\tag{13}%
\end{equation}
where $x=\frac{1}{2}\left(  1+\sqrt{1+C^{2}}\right)  $. 

The numerical simulations of entanglement of the spin states are performed for
special case $I=3/2$ using the software based on the \textit{Mathematica}
package. In the equilibrium the states of the system determined by Eq. (7) are
separable without applying external magnetic filed $\left(  \alpha=0\right)  $
at any temperature. At large temperature $\left(  \beta<1\right)  $ and low
magnetic field strength $\left(  \alpha<1\right)  $ the concurrence is zero.
Entanglement appears in the course of temperature decrease and increasing the
magnetic field.

The concurrence and the entanglement of formation dependences on the magnetic
field, quadrupole coupling constant, and temperature is qualitatively
independent of the angles. As example, we present below the concurrence as a
function of the parameters $\alpha$ and $\beta$ for the concurrence maximum
for a spin in the five-coordinated copper ion site ($\eta=0.14)$ at
$\theta=0.94\ $and $\varphi=0$ (Fig. 1). The concurrence and the entanglement
of formation increase with the magnetic field strength and inverse temperature
and reache their maximum value. Then the concurrence and the entanglement of
formation decrease with increasing the magnetic field strength (Fig. 2).\emph{
}Another dependence of the concurrence and the entanglement of formation on
temperature is observed (Fig. 3). At a high temperature concurrence and the
entanglement of formation are zero. With a decrease of temperature below a
critical value the concurrence and the entanglement of formation monotonically
increase till a limiting value. The critical temperature and limiting value
are determined by a ratio of the Zeeman and quadrupole coupling energies,
$\alpha/\beta.$

\section{\textbf{Discussion and conclusions}}

The obtained results show that the entangled states can be generated between
the states of a single nuclear spin $\frac{3}{2}$. From a point of of view of
quantum information processing the considered system is isomorphic to a system
consisting of two dipolar coupling spins $\frac{1}{2}$. The same quantum
logical gates can be realized using the both systems. Therefore the obtained
entanglement can be considered as entanglement between qubits formed by states
of a single particle. On the other hand, a single spin 3/2 isomorphic to a
system consists of two dipolar coupling spins 1/2 and entanglement between the
states of a spin 3/2 can be considered as entanglement between two effective
spins 1/2. The behavior of entanglement of a spin 3/2 is very similar to that
for the system consisting of two dipolar coupling spins $\frac{1}{2}$
\cite{Furman2010}. It was obtained that in zero magnetic field the states of
the both spin systems, two spins $\frac{1}{2}$ and spin $\frac{3}{2}$, are in
separable states. These systems become entangled with increasing the magnetic
field. Then, with a further increase of the magnetic field the spin states of
the both systems tend to a separable one.

It has recently been shown that, in a system of nuclear spins $s$ = 1/2, which
is described by the idealized XY model and dipolar coupling spin system under
the thermodynamic equilibrium conditions, entanglement appears at very low
temperatures $T\approx0.5\div0.3$
$\mu$%
K \cite{Fel'dman2007,Furman2010}. In a non-equilibrium state of the spin
system, realized by pulse radiofrequency irradiations, estimation of the
critical temperature at which entanglement appears in a system of spins
$\frac{1}{2}$ gives $T\leq0.027$ K \cite{Fel'dman2008}. The calculation for
$^{63}Cu$ in the five-coordinated copper ion site of $YBa_{2}Cu_{3}%
O_{7-\delta}$ at $\alpha/\beta=1$, $\eta=0.14$ and $eQq_{zz}=$ $62.8$ MHz,
gives that the concurrence appears at $\beta=0.6$ (Fig. 4).$\,\allowbreak$This
$\beta$ value corresponds to temperature $T\approx5$ mK. This estimated value
of critical temperature is by three orders greater than the critical
temperature estimated for the two dipolar coupling spins under the
thermodynamic equilibrium \cite{Furman2010}.

In conclusion, performing investigation has shown that entanglement can be
achieved by applying a magnetic field to a single spin 3/2 at low temperature.
Concurrence and the entanglement of formation depend on the orientation
between the external magnetic field and PAF axes. At $\theta=0$ and $\pi$ the
states are separable at any conditions. At $\eta=0.14$ the concurrence and the
entanglement of formation reache their maximum value \emph{ }at $\theta=0.94$
and $\varphi=0$ and$\ \pi$ . This effect can open a way to manipulate with the
spin states by a rotation of the magnetic field or a sample.

\bigskip

Figure Captions

\ \ 

\bigskip

\bigskip

Fig. 1 (Color online) The maximum concurrence as a function of the parameters
$\alpha$ and $\beta$ at $\eta=0.14,$ $\theta=0.94,$ $\varphi=0.$

\bigskip

Fig. 2 (Color online) Concurrence (a) and the entanglement of formation (b)
vs. magnetic field at $T$ = const for various quadrupole interaction
constants: black solid line -- $\beta=2$; red dashed line --$\beta=6$ ; green
dotted line -- $\beta=8$; blue dash-doted line -- $\beta=12$.

\bigskip

Fig. 3 Concurrence (a) and the entanglement of formation (b) as a function of
temperature at $\frac{\alpha}{\beta}=0.5$ (black solid line), $\frac{\alpha
}{\beta}=1$ (red dashed line), and $\frac{\alpha}{\beta}=2$ (blue dotted line)
at $\eta=0.14$, $\theta=0.94$, $\varphi=0$ Temperature is given in units of
$\frac{eQq_{ZZ}}{4I(2I-1)k_{B}}$.

\bigskip

\bigskip

\bigskip
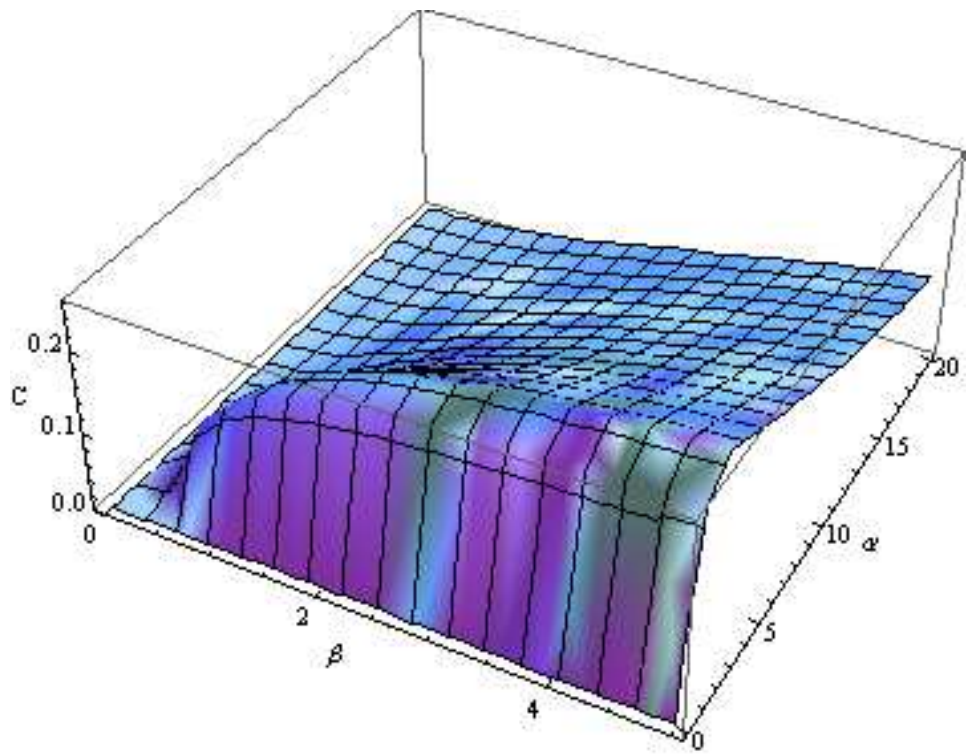
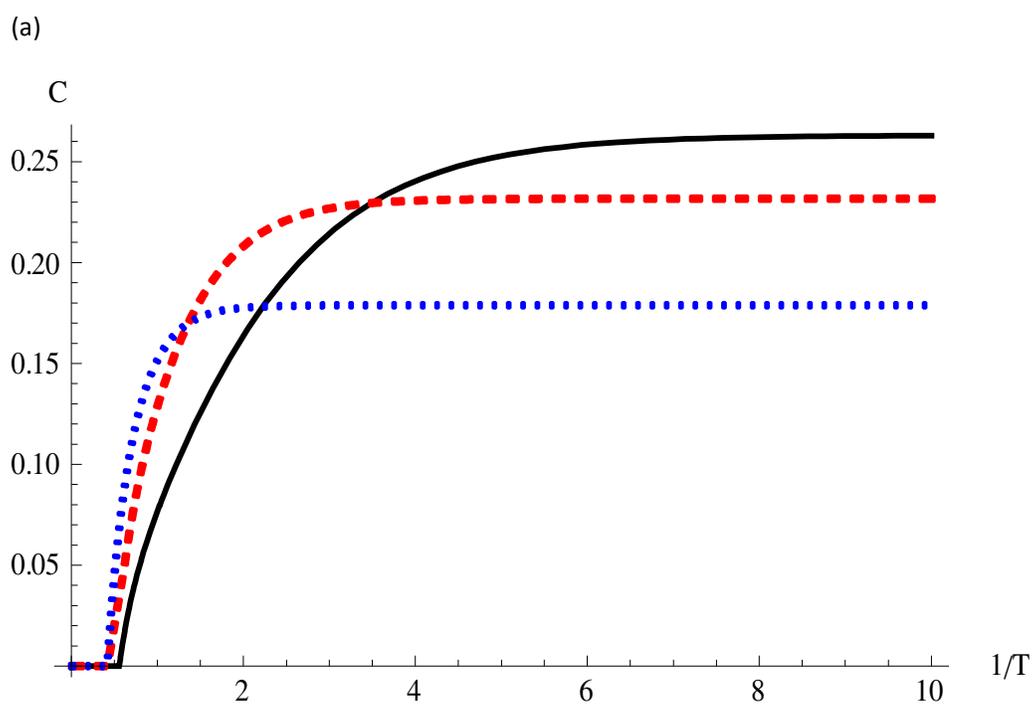
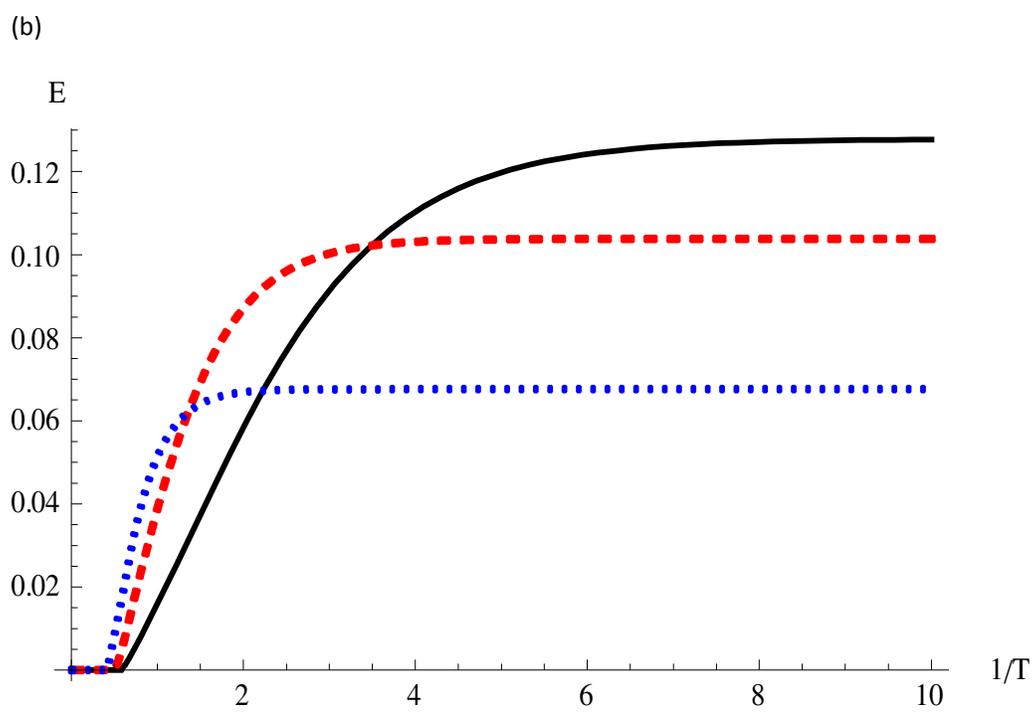
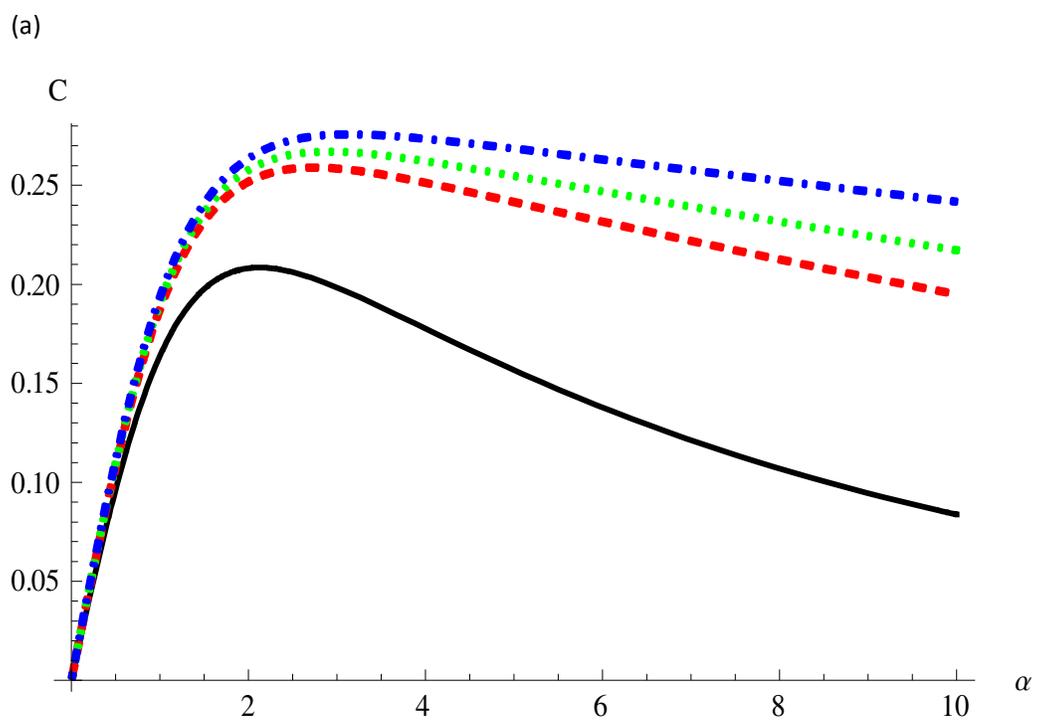
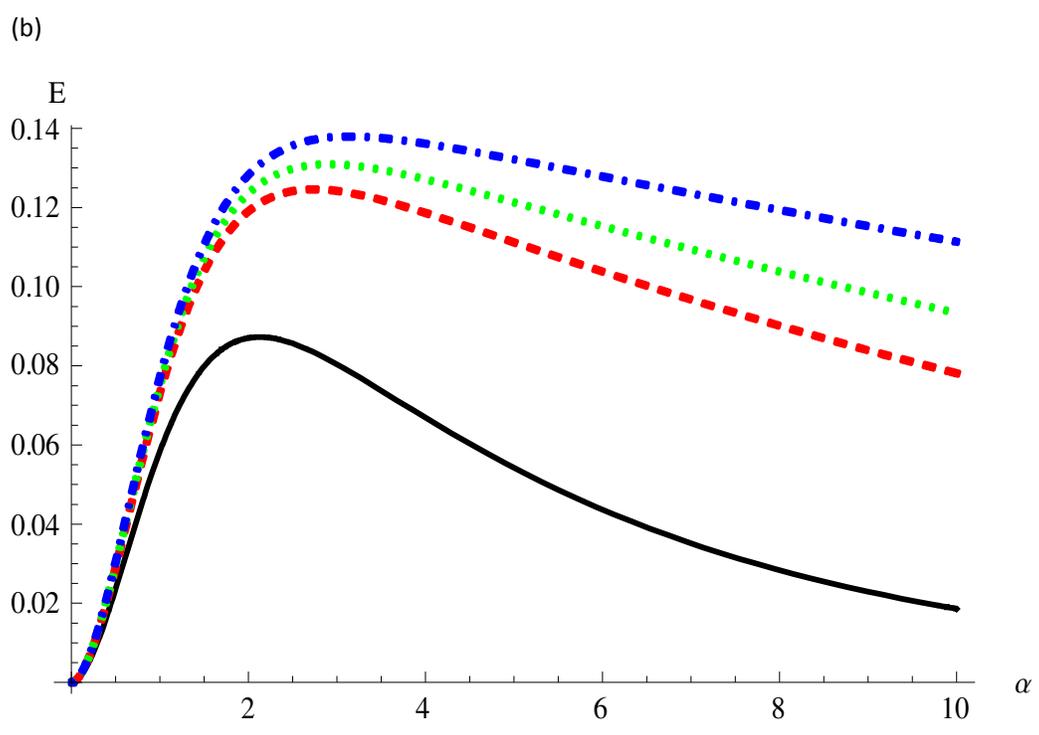
\end{document}